\documentclass[sigconf]{acmart}

\usepackage{color}
\usepackage{bbm}
\usepackage{multirow}

\usepackage[inline]{enumitem}
\usepackage{graphicx}
\usepackage{subcaption}
\usepackage{sistyle}
\usepackage[ruled]{algorithm2e}
\usepackage{booktabs}
\usepackage{caption}
\SIthousandsep{,}
\usepackage{makecell}
\usepackage{amsmath}
\usepackage[english]{babel}
\usepackage{hyperref}
\usepackage{array} 
\usepackage{graphicx}
\usepackage{algorithmic}
\usepackage{tikz}
\usepackage{wrapfig}
\usepackage{acronym}
\usepackage{graphicx} 
\usepackage{float}
\usepackage{subcaption}  
\usepackage{fontawesome}
\usepackage[most]{tcolorbox}

\newcommand{\heading}[1]{\smallskip\noindent\textbf{#1.}}

\AtBeginDocument{
  \providecommand\BibTeX{{
    \normalfont B\kern-0.5em{\scshape i\kern-0.25em b}\kern-0.8em\TeX}}}

\acrodef{CV}{computer vision}
\acrodef{IR}{information retrieval}
\acrodef{LLM}{large language model}
\acrodef{MDP}{Markov decision process}
\acrodef{NLP}{natural language processing}
\acrodef{NRM}{neural ranking model}
\acrodef{RL}{reinforcement learning}
\acrodef{RL-MARA}{Multi-grAnular Ranking Attack}
\acrodef{MoE}{mixture-of-experts}

\copyrightyear{2025}
\acmYear{2025}
\setcopyright{cc}
\setcctype{by}
\acmConference[SIGIR-AP 2025]{Proceedings of the 2025 Annual International
ACM SIGIR Conference on Research and Development in Information Retrieval in
the Asia Pacific Region}{December 7--10, 2025}{Xi'an, China}
\acmBooktitle{Proceedings of the 2025 Annual International ACM SIGIR
Conference on Research and Development in Information Retrieval in the Asia
Pacific Region (SIGIR-AP 2025), December 7--10, 2025, Xi'an,
China}
\acmDOI{10.1145/3767695.3769498}
\acmISBN{979-8-4007-2218-9/2025/12}

\makeatother


\author{Yingchen Zhang}
\orcid{0009-0003-3979-6069}
\affiliation{
 \institution{State Key Laboratory of AI Safety, Institute of Computing Technology, Chinese Academy of Sciences}
 \institution{University of Chinese Academy of Sciences}
 \city{Beijing}
 \country{China}
}
\email{zhangyingchen23s@ict.ac.cn}

\author{Ruqing Zhang}
\orcid{0000-0003-4294-2541}
\authornote{Jiafeng Guo and Ruqing Zhang are the corresponding authors.}
\affiliation{
 \institution{State Key Laboratory of AI Safety, Institute of Computing Technology, Chinese Academy of Sciences}
 \institution{University of Chinese Academy of Sciences}
 \city{Beijing}
 \country{China}
}
\email{zhangruqin@ict.ac.cn}

\author{Jiafeng Guo}
\orcid{0000-0002-9509-8674}
\authornotemark[1]
\affiliation{
 \institution{State Key Laboratory of AI Safety, Institute of Computing Technology, Chinese Academy of Sciences}
 \institution{University of Chinese Academy of Sciences}
 \city{Beijing}
 \country{China}
}
\email{guojiafeng@ict.ac.cn}

\author{Wenjun Peng}
\orcid{0000-0003-2392-4946}
\author{Sen Li}
\orcid{0000-0002-2124-953X}
\affiliation{
 \institution{Researcher}
 \city{Hangzhou}
 \country{China}
}
\email{pengwj@mail.ustc.edu.cn}
\email{lisen.lisen@alibaba-inc.com}

\author{Fuyu Lv}
\orcid{0000-0001-5918-093X}
\affiliation{
 \institution{Researcher}
 \city{Hangzhou}
 \country{China}
}
\email{fuyu.lfy@alibaba-inc.com}

\author{Xueqi Cheng}
\orcid{0000-0002-5201-8195}
\affiliation{
 \institution{State Key Laboratory of AI Safety, Institute of Computing Technology, Chinese Academy of Sciences}
 \institution{University of Chinese Academy of Sciences}
 \city{Beijing}
 \country{China}
}
\email{cxq@ict.ac.cn}
\renewcommand{\shortauthors}{Yingchen Zhang et al.}

\begin{document}

\title[C2T-ID: Converting Semantic Codebooks \\ to Textual Document Identifiers for Generative Search]{C2T-ID: Converting Semantic Codebooks \\ to Textual Document Identifiers for Generative Search}

\begin{abstract}
Designing document identifiers (docids) that carry rich semantic information while maintaining tractable search spaces is a important challenge in generative retrieval (GR). Popular codebook methods address this by building a hierarchical semantic tree and constraining generation to its child nodes, yet their numeric identifiers cannot leverage the large language model’s pretrained natural language understanding. Conversely, using text as docid provides more semantic expressivity but inflates the decoding space, making the system brittle to early‐step errors. To resolve this trade‐off, we propose C2T‐ID:(i) first construct semantic numerical docid via hierarchical clustering; (ii) then extract high‐frequency metadata keywords and iteratively replace each numeric label with its cluster’s top-$K$ keywords; and (iii) an optional two-level semantic smoothing step further enhances the fluency of C2T‐ID. Experiments on Natural Questions and Taobao’s product search demonstrate that C2T-ID significantly outperforms atomic, semantic codebook, and pure‐text docid baselines, demonstrating its effectiveness in balancing semantic expressiveness with search space constraints.
\end{abstract}

\maketitle

\section{Introduction}
\label{sec:intro}
Recently, with the rapid progress in representation learning \cite{sentbert,Contrastive_Learning} and pre-trained language models \cite{bart,gpt,llama}, neural retrieval models \cite{dpr,colbert,LargeDualEncoders,MultiVectorRetrievalSparse,dsi} have led to exciting breakthroughs in information retrieval (IR). 
Traditional IR paradigms, such as term-based retrieval \cite{bm25} and dense retrieval \cite{dpr,colbert} typically build an index over the corpus and then return relevant documents by measuring similarity between the query and the index. 
In recent years, driven by the rapid advancement of large language models (LLMs) \cite{gpt,llama,deepseek-r1}, generative retrieval (GR) \cite{dsi,gere,nci,seal} has emerged as a new end-to-end retrieval paradigm. 
GR directly leverages an autoregressive LLM to generate document identifiers (docids) token by token given a query. 
GR has achieved competitive performance compared to traditional IR methods on tasks such as document retrieval \cite{ripor,pag}, code retrieval \cite{codedsi}, and book retrieval \cite{grbook}.

\heading{Docid design}  
In GR, the autoregressive generation of docids is brittle: an incorrect prediction at any step will result in the target document being omitted from the final retrieval results \cite{dsi,seal,nci,dsi++}. 
Consequently, designing suitable docid for candidate documents presents a significant challenge. 
Existing methods can be broadly classified into three categories:
\begin{enumerate*}[label=(\roman*)]
    \item  \textbf{Atomic Numeric Docid \cite{dsi,gr-dr}}: Each document is assigned a unique integer identifier, which the model generates in one shot. This approach carry no semantic information and are rarely used today;
    \item \textbf{Semantic Codebook Docid \cite{continual-gr,ultron,dsi++,learn-to-tokenize-gr,nci,ripor}}: A discrete codebook is first constructed via hierarchical clustering \cite{dsi,nci} or product quantization \cite{continual-gr}. Each document is represented by a sequence of codebooks, which the model decodes step by step under a trie constraint. This injects limited semantic structure while preserving hierarchical search constraints, but the outputs remain numeric, making it difficult to fully leverage the LLM’s pretrained capability; and
    \item  \textbf{Textual docid \cite{gere,corpus-brain,non-parametric-gr,ultron,seal,learn-to-tokenize-gr,minder,tome}}: Documents are identified by actual natural language text (such as titles \cite{gere,corpus-brain,non-parametric-gr}, URLs \cite{ultron}, or n-gram set \cite{seal,learn-to-tokenize-gr,minder,tome}) to maximize the use of a pretrained LLM’s semantic understanding. However, the candidate space at each decoding step expands to the entire vocabulary, increasing the likelihood of early-generation errors.
\end{enumerate*}

\begin{figure}[t]
    \centering
    \includegraphics[width=\linewidth]{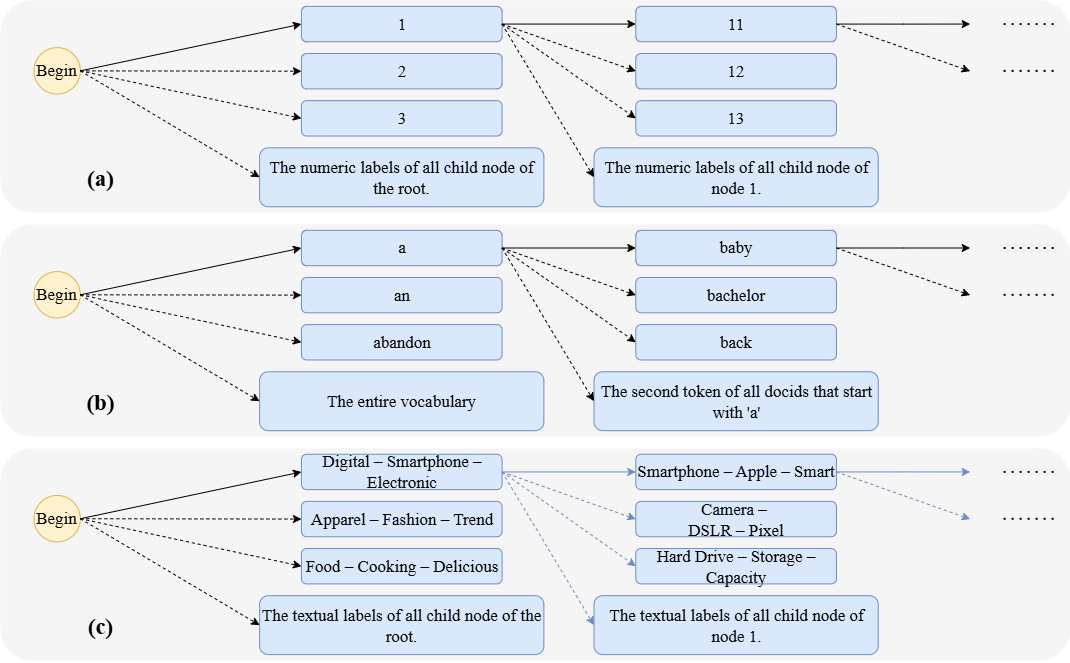}

    \caption{Comparison of C2T-ID with other docid methods. (a) is an example of the decoding for semantic codebook docid, (b) for textual docid, and (c) for our C2T-ID.}
    \label{figure:overview}
\end{figure}  

The latter two approaches represent the current mainstream in docid design, differing in how they leverage document priors:  semantic codebook docid preserves hierarchical search constraints but offers limited semantic expressiveness, while textual docid delivers rich semantics at the cost of an inflated search space.
We aim to develop a new docid construction method that balances these two, carrying rich textual priors while preserving hierarchical search space constraints.

\heading{Our method}  
In this paper, we propose \textbf{C2T-ID}, which converts semantic codebook paths into structured textual document identifiers. 
Specifically, 
\begin{enumerate*}[label=(\roman*)]
    \item we first build a hierarchical clustering tree over the document collection and assign each document a numeric path as a coarse semantic numerical docid;
    \item we then extract high-frequency keywords from each document’s metadata (e.g., Wikipedia categories or e-commerce attributes);
    \item finally, we replace each numeric label with the top-\(K\) keywords of its cluster, joined by hyphens to form the C2T-ID; and
    \item as an optional enhancement, we apply a two-level semantic smoothing: an LLM rewrites intra-cluster phrases and inserts natural connective words between hierarchy levels to further improve the fluency of the C2T-ID while preserving the tree structure.
\end{enumerate*}
The comparison between C2T-ID and common docid methods is shown in Figure \ref{figure:overview}.

\heading{Experiments}  
To evaluate C2T-ID, we conduct experiments on two datasets: Natural Questions (NQ) \cite{nq} and Tobao’s product search. 
We compare GR performance using C2T-ID against baselines with atomic docid \cite{dsi}, semantic codebook docid \cite{dsi}, and textual docid using title \cite{gere}. 
Our results demonstrate the effectiveness of C2T-ID.

\section{Related works on Genetative Retrieval (GR)}
\label{sec:related_works}

GR is an emerging paradigm that leverages autoregressive LLMs for end-to-end IR \cite{dsi,gere,nci,seal}. 
In traditional retrieval paradigms (such as term-based retrieval\cite{bm25} or dense retrieval \cite{dpr,colbert}), the IR model first builds an index over the corpus (inverted lists or dense embeddings) and then returns relevant documents by measuring query–index similarity (lexical matching or vector dot products). 
In contrast, GR aims to integrate all necessary corpus information into a single, unified LLM. Specifically, it trains the model using standard maximum likelihood estimation on document–docid pairs and query–docid pairs, thereby teaching the model both the content of documents and the meaning of their identifiers. 
At inference time, GR directly decodes document identifiers from the constrained docid space conditioned for the input query.

Research on GR can be broadly categorized into three areas. 
Specifically,
\begin{enumerate*}[label=(\roman*)]
    \item a line of work has focused on constructing docid: early approaches employed purely numeric docid \cite{dsi,gr-dr}, while more recent methods have adopted semantically enriched numeric docid (codebooks) \cite{continual-gr,ultron,dsi++,learn-to-tokenize-gr,nci,ripor} or even directly used document-related text, such as titles \cite{gere,corpus-brain,non-parametric-gr}, URLs \cite{ultron} and n-gram set \cite{seal,learn-to-tokenize-gr,minder,tome}, as docid;
    \item some works seek to enhance GR training by leveraging large-scale multi-task learning \cite{ultron} and incorporating ranking loss \cite{listwise-gr,ripor,ltgtr}, or addressing the limitations of training data through data augmentation and alignment techniques \cite{ultron,dsi-qg,nci}; and
    \item a growing body of work has explored applying GR to specific IR domains such as code retrieval \cite{codedsi}, book retrieval \cite{grbook} and image retrieval \cite{cross-modal-gr,image-gr}, or to downstream tasks including entity linking \cite{genre} and fact verification \cite{gere}.
\end{enumerate*}

\section{Method}
\label{sec:method}
\subsection{Overview}

Our C2T-ID converts semantic codebooks into structured, multi-token textual docids that enrich GR with document priors while preserving the clustering tree constraints. 
Specifically, 
\begin{enumerate*}[label=(\roman*)]
    \item we first build a hierarchical clustering tree over the corpus and assign each document a numeric path as a semantic numerical docid (Section~\ref{subsec:semantic_codebooks_construction});
    \item we then extract a set of high-frequency keywords from each document (Section~\ref{subsec:document_prior_information_extraction});
    \item finally, we replace each numeric label with the hyphen-joined string of its cluster’s top-$k$ keywords to form the textual C2T-ID (Section~\ref{subsec:textual_label_replavement}); and
    \item as an optional enhancement, we apply a two-level semantic smoothing to improve the fluency of C2T-ID while maintaining the original tree structure (Section~\ref{subsec:semantic_smoothing}).
\end{enumerate*}

This design allows each docid to carry rich textual signals that GR can readily interpret, fully leveraging pre-trained LLM's capability. 
Moreover, because each generation step follows the fixed clustering tree, the search space at each decoding step remains confined to a limited set of nodes (rather than the entire vocabulary). 
Thus, C2T-ID strikes a balance between semantic expressiveness and constrained generation.

\subsection{Semantic codebooks construction}
\label{subsec:semantic_codebooks_construction}
Following the general approach for constructing codebooks as semantic numerical docid \cite{dsi,nci}, we first generate a vector representation for each document, then apply hierarchical k-means clustering on corpus $D = \{d_1, \dots, d_N\}$ to build a tree structure $T = (V, E)$. 
Specifically, starting from the root node $r_0$, we partition all documents into at most $k$ clusters; for any cluster containing more than $c$ documents, we recursively apply $k$-means until every leaf cluster has at most $c$ documents. 
At the $i$-th level of the tree, each node $v \in V$ is assigned an integer label $r_i \in [0, k)$ (for leaf nodes, $r_i \in [0, c)$).
Each document $d_j$ corresponds to a unique routing path $\ell_j = (r_0, r_1, \dots, r_m)$ from the root, and this sequence of indices forms its semantic numerical docid (also called the codebook).

\subsection{Document prior information extraction}
\label{subsec:document_prior_information_extraction}
We further aim to endow each docid with finer-grained semantic signals so that LLMs can exploit their understanding and reasoning over text. 
In general, most candidate entities or documents come with external knowledge sources or metadata from which we can harvest prior information (e.g., category labels, keywords, and attribute fields). 
By injecting these priors into the docid, we can provide additional semantic cues that help the GR accurately focus on the relevant documents. We use distinct extraction strategies for two representative scenarios: 
\begin{itemize}[noitemsep, left=0pt]
    \item  \textbf{Wikipedia Document Retrieval}: Wikipedia maintains a taxonomy of document categories. Each article is assigned to one or more category nodes upon inclusion. We retrieve the full list of categories for each article via the MediaWiki API, filter out overly general classes such as ``All pages'' or ``Uncategorized'', and thus obtain a concise set of text labels that highly summarize the document’s topic.

    \item \textbf{Taobao Product Retrieval}: On the Taobao platform, each product is annotated with rich metadata, including its title, description, brand, and various attribute fields. We tokenize and remove stop words from these text fields, then aggregate keywords from attribute fields (e.g., ``material'', ``usage'',``style''), selecting the most frequent terms as the product’s semantic priors.

\end{itemize}

At this point, we construct a set of keywords for each candidate document to capture its core topics and attributes. 
Although our implementations are dataset-specific, the general principle of extracting high-quality priors from available metadata can be readily applied to other corpora.

\vspace{-6pt}

\subsection{Textual label replacement}
\label{subsec:textual_label_replavement}
Each document $d_j$ is associated with a numeric path $\ell_j = (r_0, r_1, \dots,\\ r_m)$ in the clustering tree described in Section \ref{subsec:semantic_codebooks_construction}. 
To convert these pure numeric labels into semantically rich textual identifiers, we first aggregate all keywords extracted in Section \ref{subsec:document_prior_information_extraction} from the documents under each node $v$ and count their frequencies. 
We then sort these keywords in descending order of frequency and select the top-$K$ as that cluster’s label. 
Within each cluster, the top-$k$ keywords are concatenated into a single substring using a hyphen (``–'') in the same order. 
Finally, for document $d_j$, we replace each numeric label $r_i$ in its path with the corresponding hyphen-joined keyword substring for cluster $r_i$, and then concatenate these substrings across levels—again using ``–'' as the separator:
\begin{equation}
C2T-ID
= \bigl[\mathrm{Top}_K(r_0)\bigr]\;\Vert\;\bigl[\mathrm{Top}_K(r_1)\bigr]\;\Vert\;\dots\;\Vert\;\bigl[\mathrm{Top}_K(r_m)\bigr],    
\end{equation}
where $\bigl[\mathrm{Top}_K(r_i)\bigr]$ denotes the hyphen-joined string of the $K$ highest-frequency keywords for node $r_i$. This hyphen-delimited sequence C2T-ID preserves the original hierarchical structure while injecting clear semantic cues, providing GR models with concise, structured prompts during decoding.

\vspace{-6pt}

\subsection{Semantic smoothing (Optional)}
\label{subsec:semantic_smoothing}
Although simple hyphen-separated keywords preserve structure, they can lead to awkward, semantically disjoint docids. 
For example, once the GR has decoded ``phone-'' as part of a Textual docid, it will almost always predict ``case'' next, because ``phone–case'' is common, even if the user’s query is about ``phone battery'', not ``phone-case''. 
To mitigate this issue, we introduce semantic smoothing as an optional enhancement. 
To ensure the original hierarchical structure remains enforced after smoothing, and to avoid excessively expanding the search space, we apply a two-level smoothing strategy. 
Specifically,
\begin{enumerate*}[label=(\roman*)]
    \item within each cluster’s $\mathrm{Top}_K$ keywords, we invoke a LLM to paraphrase and reorder the words into a short, natural phrase. 
    For instance, instead of ``phone–battery–charger'', the LLM will produce ``battery charger for phone''. Because GR must complete all tokens under the current node before moving on, reordering does not introduce new branch points in the decoding trie; and
    \item between adjacent cluster levels, we replace the rigid hyphen ``–'' with a brief connective word generated by the LLM (e.g., “with,” “of,” “for”). Crucially, we fix the sequence of clusters so that once the decoder begin to generate the child cluster’s span, it cannot stray into unrelated text.
\end{enumerate*}

By constraining the LLM’s rewrites to keywords within each node and preserving the parent–child order between nodes, this semantic smoothing introduces natural fluency without altering the GR search space at any generation step. Since this step incurs substantial LLM invocation overhead and offers only limited performance gains, we regard it as an optional enhancement.

\begin{table*}[t]
  \centering
  \caption{Comparison of retrieval performance on NQ and Taobao. The best results are shown in bold, the second-best are underlined, the best GR result is marked with $\dagger$, the second-best GR result is marked with $\ddagger$, and inapplicable entries are indicated by “–”. We implement standard GR using four different docids (first two like DSI, the third like GERE, and ours). The results demonstrate that our C2T-ID yields substantially improve GR performance than the alternatives. Moreover, even with a basic GR implementation, our method outperforms traditional IR on the large-scale, complex Taobao product-search task, suggesting significant potential as task complexity, dataset size, and model scale increase.
}
  \label{tab:main_results}
  \resizebox{0.8\textwidth}{!}{
  \begin{tabular}{lccccc}
    \toprule
    \multirow{2}{*}{\textbf{Method}} 
      & \multicolumn{3}{c}{\textbf{NQ}} 
      & \multicolumn{2}{c}{\textbf{TB}} \\
    \cmidrule(r){2-4} \cmidrule(l){5-6}
      & \textbf{Hits@5} & \textbf{Hits@20} & \textbf{MRR@20}
      & \textbf{Hits@5} & \textbf{Hits@20} \\
    \midrule
    \multicolumn{6}{l}{\textit{Term-based and dense retrieval}} \\
    BM25
      & \underline{43.6} & \underline{62.9} & \underline{34.6}
      & 42.8 & 55.1 \\
    DPR
      & \textbf{63.8} & \textbf{80.1} & \textbf{47.3}
      & \underline{62.8} & \underline{73.1} \\
    \midrule
    \multicolumn{6}{l}{\textit{Generative retrieval with different DocIDs trained using a DSI-style workflow}} \\
    Atomic DocID
      & 12.3 & 22.7 & 16.9
      & 5.7  & 7.2 \\
    Semantic Codebook DocID
      & $28.3^\ddagger$ & $47.3^\ddagger$ & 26.3
      & $51.6^\ddagger$ & $58.0^\ddagger$ \\
    Textual DocID (Title)
      & 25.5 & 43.3 & $27.1^\ddagger$
      & –    & –    \\
    \midrule
    \multicolumn{6}{l}{\textit{Ours}} \\
    \textbf{C2T-ID}
      & $34.3^\dagger$ & $51.5^\dagger$ & $31.2^\dagger$
      & $\textbf{64.0}^\dagger$ & $\textbf{77.5}^\dagger$ \\
    \bottomrule
  \end{tabular}
  }
\end{table*}

\section{Experiments}
\label{sec:experiments}

\subsection{Experimental settings}
\heading{Datasets and Metrics}
We evaluate our C2T-ID on the following two datasets:
\begin{enumerate*}[label=(\roman*)]
    \item \textbf{Natural Questions (NQ) \cite{nq}}: a Wikipedia‐based question answering retrieval task, where the candidate documents are Wikipedia articles and the queries are natural language questions; and
    \item \textbf{Taobao’s Product Search (TB)}: an e-commerce product retrieval dataset, with a candidate set of millions of Taobao’s items and queries drawn from real user search logs. TB consists of product clusters under the 3C category in Taobao Mall obtained via pre‐clustering, together with 2.5 million query–cluster pairs used for training.
\end{enumerate*}

The evaluation metrics include \textbf{Hits} and \textbf{MRR}, which measure the hit rate within the top-$K$ results and the overall ranking quality respectively.

\heading{Baselines}
We implement GR baselines based on three different docids:
\begin{enumerate*}[label=(\roman*)]
    \item \textbf{Atomic Docid}: Same to the Unstructured Atomic Identifiers in DSI \cite{dsi}, each document is assigned a unique integer identifier that is generated in a single step;
    \item \textbf{Semantic Codebook Docid}: Similar to the Semantically Structured Identifiers in DSI \cite{dsi}, we first build a hierarchical clustering codebook and then generate the numeric codebook layer by layer; and
    \item \textbf{Textual Docid}: We directly use the document title as the docid, following the approach used in GERE \cite{gere}.
\end{enumerate*}
In addition, we also compare against two classic IR methods outside of GR: \textbf{BM25} \cite{bm25} and \textbf{DPR} \cite{dpr}.

\heading{Implementation details}
For the NQ dataset, we use BART-large \cite{bart} as the backbone.
For the TB dataset, we use LLAMA-3.1-8B \cite{llama3}as the backbone. 
We follow the standard GR training procedure \cite{dsi} and use the Adam optimizer \cite{adam} with its default settings to train the model for both indexing and retrieval.
All training and inference are performed on two NVIDIA H20 GPUs.
We set the hierarchical clustering parameters to $k=30$ and $c=30$, and use $K=3$ as the number of top frequency keywords for textual replacement.

\subsection{Experimental results}

\heading{Main results}
Table \ref{tab:main_results} presents the comparison results between C2T-ID and baseline methods on the NQ and TB datasets. We can draw the following conclusions:
\begin{enumerate*}[label=(\roman*)]
    \item C2T-ID significantly outperforms GR models using atomic docid, semantic codebook docid, and pure textual docid on all evaluation metrics across both datasets;
    \item although we employ only a basic GR implementation and overall performance still lags behind conventional IR methods, C2T-ID already matches or exceeds them on certain metrics. This suggests that integrating more advanced GR architectures could enable GR to fully overtake traditional retrieval methods;
    \item purely using document titles as textual docid underperforms semantic codebook docid, likely due to the vastly expanded search space. By contrast, C2T-ID injects rich textual priors while retaining hierarchical decoding constraints, yielding better results than the codebook approach on every metric; and
    \item C2T-ID delivers especially pronounced gains on the more challenging Taobao product retrieval task. Given our use of a larger model for this task, this finding indicates that as task complexity, dataset size, and model scale increase, C2T-ID may realize even greater benefits.
\end{enumerate*}
Overall, C2T-ID effectively balances semantic expressiveness and hierarchical constraints, offering a practical and efficient new direction for docid design in GR.

\heading{Semantic smoothing experiment}
To evaluate the impact of the optional two-level semantic smoothing, we conduct an ablation study on NQ dataset. 
Specifically, we compare the performance of C2T-ID with and without the smoothing module. 

\begin{figure}[h]
    \centering
    \includegraphics[width=\linewidth]{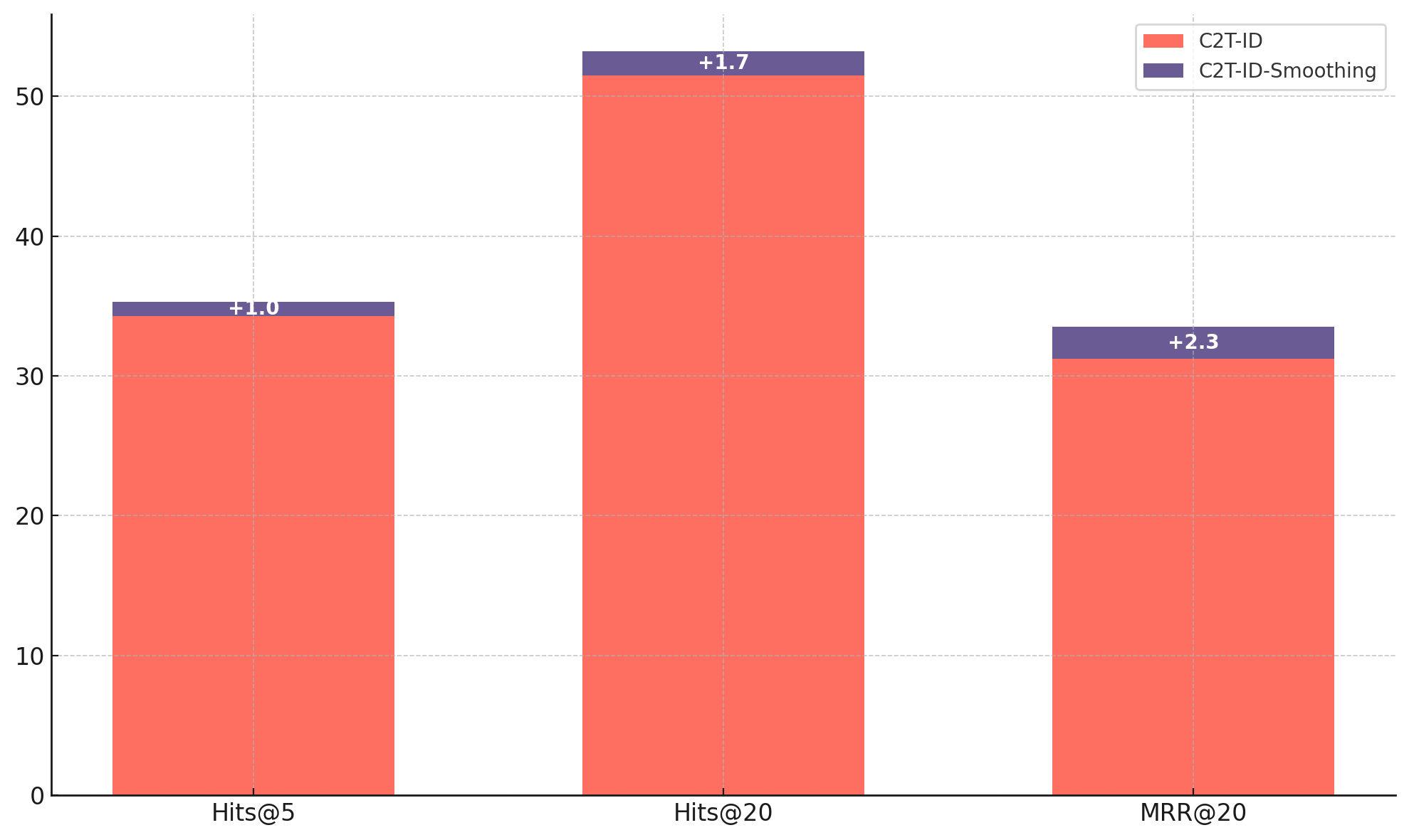}

    \caption{Results of two-level semantic smoothing experiment on the NQ dataset.}
    \label{figure:smoothing_results}
\end{figure}

Results are presented in Figure~\ref{figure:smoothing_results}.

The results indicate that incorporating semantic smoothing delivers consistent but modest improvements across all metrics. 
Considering the substantial overhead introduced by LLM-based smoothing, this step can be omitted in cost-sensitive settings without compromising the overall performance of C2T-ID.

\heading{Zero-shot experiment}
In the zero-shot experiment, we train each GR model solely on document–docid pairs, without any query–docid supervision \cite{dsi}. 
This means the models learn to associate documents with their identifiers but never see real query examples during training. 
At test time, they must decode docids directly from unseen queries, simulating a true zero-shot retrieval scenario where the model has no prior exposure to query-document mappings.

\begin{table}[h]
  \centering
  \caption{Zero-shot retrieval performance on NQ.}
  \label{tab:zeroshot_results}
  \begin{tabular}{lccc}
    \toprule
    \textbf{Method} & \textbf{Hits@5} & \textbf{Hits@20} & \textbf{MRR@20} \\
    \midrule
    Atomic Docid             & 9.3 & 18.5 & 12.4 \\
    Semantic Codebook Docid  & 22.0 & 38.6 & 15.5 \\
    Textual Docid (Title)    & \underline{22.2} & \underline{41.1} & \underline{25.3} \\
    \midrule
    \textbf{C2T-ID (ours)}  & \textbf{33.5} & \textbf{52.3} & \textbf{30.3}\\
    \bottomrule
  \end{tabular}
\end{table}

As shown in Table \ref{tab:zeroshot_results}, C2T-ID achieves the highest retrieval performance among all docid methods. 
Remarkably, its zero-shot accuracy is very close to its fully supervised counterpart, indicating almost no degradation despite the absence of query-level training. 
Other text-based docid methods also exhibit only minor drops compared to their supervised versions, underscoring the importance of richly semantic docid.
\vspace{-5pt}
\section{Conclusion and limitations}
\label{sec:conclusion}
In this work, we propose C2T-ID, a novel method that converts semantic codebook docid into structured textual docid. 
Specifically, 
\begin{enumerate*}[label=(\roman*)]
    \item we first generate coarse-grained numerical docids via hierarchical clustering;
    \item then we iteratively replace each level’s numeric labels with high-frequency keywords extracted from document metadata; and 
    \item optionally applying a two-level semantic smoothing, C2T-ID injects rich textual priors while preserving the original clustering-tree constraints. 
\end{enumerate*}
Empirical results show that GR with C2T-ID achieves significant retrieval performance improvements on both the Natural Questions and Taobao’s product search datasets compared to atomic docid, semantic codebook docid, and textual docid using title.

Despite its balance between semantic expressiveness and search space constraints, C2T-ID has several limitations:
\begin{enumerate*}[label=(\roman*)]
    \item C2T-ID typically uses more tokens than other docid construction approaches, which substantially increases inference overhead and may be unsuitable for latency-sensitive IR systems;
    \item its keyword extraction relies on dataset-specific metadata sources, requiring customized prior designs for different domains; and 
    \item the semantic smoothing introduces additional LLM calls, incurring significant docid construction costs.
\end{enumerate*}

\section{Acknowledgments}
This work was funded by the Strategic Priority Research Program of the CAS under Grants No. XDB0680102, the National Natural Science Foundation of China (NSFC) under Grants No. 62472408 and 62441229, the National Key Research and Development Program of China under Grants No. 2023YFA1011602. 
All content represents the opinion of the authors, which is not necessarily shared or endorsed by their respective employers and/or sponsors. 

\bibliographystyle{ACM-Reference-Format}
\balance
\bibliography{references}

\end{document}